\documentstyle[multicol,aps,epsfig,prl]{revtex}
\tighten
\begin{document}
\draft

\title{Resonant tunneling of interacting electrons in a
  one-dimensional wire}
\author{Yu.V Nazarov$^{1}$ and L.I. Glazman$^{2}$} 
\address{{}$^1$Department of Nanoscience, Delft University of Technology,
2628 CJ Delft, the Nethrelands}
\address{{}$^2$ Theoretical Physics
  Institute, University of Minnesota, Minneapolis, MN 55455}
  \maketitle

\begin{abstract}
We consider the conductance of a one-dimensional wire interrupted by a
double-barrier structure allowing for a resonant level. Using the
electron-electron interaction strength as a small parameter, we are
able to build a non-perturbative analytical theory of the conductance
valid in a broad region of temperatures and for a variety of the
barrier parameters. We find that the conductance may have a
non-monotonic crossover dependence on temperature, specific for a
resonant tunneling in an interacting electron system.
\end{abstract} 

\begin{multicols}{2} 
The phenomenon of resonant tunneling is well-known in the context of
electron transport physics~\cite{resonant}. The hybridization of a
discrete state localized in the barrier with the extended states
outside the barrier may strongly enhance the transmission
coefficient for electrons incident on the barrier with energy
matching the energy of the localized state. For a single electron,
the transmission coefficient at energies close to the resonance is
given by the Breit-Wigner formula~\cite{resonant}. However, if the barrier
carrying the resonant level separates conductors which in
equilibrium have a finite density of mobile electrons, the problem
of resonant tunneling becomes more complex due to the
electron-electron interaction.  Manifestation of resonant tunneling
in the conductance of a solid-state device is inevitably sensitive
to this interaction.

Some of the effects of electron-electron interaction do not depend on
the dimensionality $d$ of the conductors--leads separated by the
barrier. For instance, the on-site repulsion together with the
hybridization of the localized state with the states of continua lead
to the Kondo effect in the transmission across the
barrier~\cite{resonant,Kd} at any $d$. The Fermi-edge singularity also
strongly affects the resonant tunneling~\cite{FEt,FEe} in any
dimension. The electron-electron interaction within the leads,
however, does not have a strong effect if $d>1$, and if the leads are
not disordered. In contrast, resonant tunneling across a barrier
interrupting a one-dimensional (1d) wire is modified drastically by
the interaction within the wire.
The importance of such a setting is emphasized by the recent transport
experiments with nanotubes and nanowires containing a quantum
dot~\cite{Auslaender,Dekker}.

The electron-electron interaction enhances the backscattering off the
barrier for electrons with energy close to the Fermi level~\cite{KF1}.
We find that if the discrete level is not perfectly aligned with the
Fermi level in the leads, or the barrier structure has even slight
geometrical asymmetry, then the low-temperature linear conductance
decreases to zero with the temperature, $G(T\to 0)\to 0$.  At
sufficiently high electron energies (measured from the Fermi level)
the enhancement of backscattering due to the interaction should get
weaker, and the conventional behavior of resonant tunneling prescribed
by Breit-Wigner formula may be restored. Consequently, the conductance
$G(T)$ may increase with the temperature being lowered.  How to match
these two opposite tendencies? We answer this question below by
finding the proper non-monotonic crossover function $G(T)$ for an
arbitrary asymmetry of the barrier and arbitrary position of the
resonant level with respect to the Fermi level, in the limit of weak
interaction.  The asymptotes of $G(T)$ agree with those found in
Ref.~\cite{KF,KF1} in the context of the Luttinger liquid theory.  The
universality of latter results was recently doubted in the theoretical
part of \cite{Dekker}.  We find no ground for such doubts.

We will use an analogue of the renormalization method developed in
\cite{MatveevGlazman}.  Within this method, the complicated picture of
many-electron transport is considered within the traditional
Landauer-B\"uttiker elastic scattering formalism. The role of the
interaction is to renormalize the elastic scattering amplitudes.  The
renormalization brings about an extra energy dependence of these
amplitudes. It was shown in~\cite{MatveevGlazman} that in the limit of
weak interaction the most divergent terms in perturbation theory
indeed correspond to the purely elastic processes, thus justifying the
method.  The advantage of the method is that it allows one to
investigate quantitatively the crossover between the limits of weak
tunneling and full transmission across the barrier. 

The original formulation of the method~\cite{MatveevGlazman}
disregarded the energy dependence of scattering amplitudes in the
absence of interaction.  While valid for a generic case of a single
scatterer, this assumption obviously fails to describe the resonant
tunneling.  To circumvent this, we extend the method to arbitrary
energy dependence of scattering amplitudes. First step is to derive
the first-order interaction correction to scattering amplitudes.  This
can be readily done along the lines of Ref.\cite{MatveevGlazman}.  The
correction to transmission amplitude reads
\begin{equation}
\delta t(\epsilon) = \frac{t(\epsilon)}{2} 
\int_{-\infty}^{0} \frac{d \epsilon'} {\epsilon'-\epsilon} 
[\alpha_L r_L({\epsilon}) r^*_L({\epsilon'}) + 
\alpha_R r^*_R({\epsilon'}) r_R({\epsilon})].
\label{firstorder}
\end{equation}
Here the $r_{L(R)}$ are the reflection amplitudes for electrons
incoming from the left (right), and the coefficients $\alpha_{L(R)}$
represent the interaction within the left(right) part of the 1d wire;
energies $\epsilon$ and $\epsilon'$ are measured from the Fermi level.
Transmission and reflection amplitudes $r_{L,R}$ satisfy the unitarity
relation: $r_R t^* = -r^*_L t$.  The coefficients $\alpha$ are related
to the Fourier components $V(k)$ of the corresponding
electron-electron interaction potential by $\alpha =
(V(0)-V(2k_F))/2\pi v_F)$. In the limit of weak interaction, these
coefficients determine the exponents in the edge density of
states~\cite{KF1} for each part of the channel, $\nu(\epsilon) \propto
\epsilon^{\alpha}$.

The integration over $\epsilon'$ in the first-order
correction~Eq.~(\ref{firstorder}), in general, yields a logarithmic
divergence at $\epsilon\to 0$.  This indicates that the perturbation
series in the interaction potential can be re-summed with the
renormalization method.  To account for the most divergent term in
each order of the perturbation theory in $\alpha$, we proceed with the
renormalization in a usual way~\cite{renormalization}.  On each step
of the renormalization, we concentrate on the electron states in a
narrow energy strip around $-E$, with $E>0$ being the running cut-off.
We evaluate the interaction correction due to the electrons in these
states to the scattering amplitudes at energies $\epsilon$ close to
Fermi level, $|\epsilon|<E$. These amplitudes are thus functions of
both $\epsilon$ and $E$.  We correct those amplitudes
according to Eq. \ref{firstorder}, reduce the running cut-off by the
width of the energy strip, and repeat the procedure.  This yields the
following renormalization equation:
\begin{eqnarray}
\frac{\partial t(\epsilon,E)}{\partial \ln E} =
\frac{t(\epsilon,E)}{2}\left[
\alpha_L r_L({\epsilon},E) r^*_L(-E)+ \right. \nonumber \\
 + 
\left. \alpha_R r^*_R(-E) r_R({\epsilon},E)
\right],
\label{transamp}
\end{eqnarray}
provided that $|\epsilon| < E$. We abbreviate here $r(\epsilon) \equiv
r(\epsilon,|\epsilon|)$ (and similar for $t$) indicating that the
renormalization of scattering amplitudes stops when the running
cut-off approaches $|\epsilon|$. The initial conditions for this
differential equation are set at upper cut-off energy $\Lambda$.  If
the $\epsilon$--dependence of the transmission amplitude in the absence
of interaction, $t(\epsilon,\Lambda)$, can be disregarded, then all
the energy dependence of renormalized amplitudes comes about as a
result of the renormalization procedure. The corresponding
simplification of Eq.~(\ref{transamp}) then reads
\begin{equation}
\frac{\partial |t(\epsilon)|^2}{ \partial \ln \epsilon} 
= (\alpha_R+\alpha_L) |t(\epsilon)|^2 (1-|t(\epsilon)|^2),
\label{diffform}
\end{equation}
and contains the transmission probabilities only.  This coincides with
the results of Ref.~\cite{MatveevGlazman}. However, the above
simplification is not possible in the more general case we consider
here.  One can not even deal with a single equation: the equation
(\ref{transamp}) shall be supplemented with a similar equation for one
of the reflection amplitudes,

\begin{eqnarray}
&&\frac{\partial r_L(\epsilon,E)}
{\partial \ln E}  = \frac{1}{2} \{\alpha_L [- r_L(-E) +r^2_L(\epsilon,E)r^*_L({-E})] \label{ramp}\\
&& 
+ \alpha_R r^*_R(-E) t^2({\epsilon},E)\}.
\nonumber
\end{eqnarray}

To describe resonant tunneling, we consider a compound scatterer made
of two tunnel barriers with tunnel amplitudes $t_{1,2} \ll 1$
separated by a distance $\pi v_F/\delta$. This gives rise to a system
of equidistant transmission resonances separated by energy $\delta$.
We assume that one of the resonances is anomalously close to Fermi
energy and concentrate on this one disregarding the others.  The
scattering amplitudes in the absence of interaction are then given by
common Breit-Wigner relations:
\begin{eqnarray}
&&t (\epsilon,\Lambda) =
\frac{ i \sqrt{\Gamma_L \Gamma_R}}
{(\Gamma_L+\Gamma_R)/2 -i(\epsilon - \Delta)}, \nonumber\\
&&r_L(\epsilon,\Lambda)= 
\frac{(-\Gamma_L+\Gamma_R)/2 -i(\epsilon - \Delta)}
{(\Gamma_L+\Gamma_R)/2 -i(\epsilon - \Delta)},\nonumber
\end{eqnarray}
where $\Gamma_{L,R} = |t_{1,2}|^2 \delta/2\pi$ are the level
widths with respect to the electron decay into the left(right) lead
and $\Delta$ is the energy shift of the resonance with respect to the
Fermi Level; we assume here $\Delta\ll\delta$. We disregard
possible energy dependence of $t_{1,2}$ that could be relevant
at higher energies, which allows us to take the upper cut-off $\Lambda$
to be of the order of $\delta$. The corresponding transmission
probability before the renormalizations,
\[
|t(\epsilon,\Lambda)|^2 =\frac{\Gamma_L \Gamma_R}
{(\epsilon - \Delta)^2 +(\Gamma_L +\Gamma_R)^2/4},
\]
is the usual Lorentzian function of energy.
The interaction corrections to $\Delta$ and $\Gamma_{L,R}$ which come
from bigger energy scales, $\delta < E <E_F$, are assumed to be
included in the definitions of these quantities.

The next step is to solve the renormalization
equations~(\ref{transamp}) and (\ref{ramp}). To stay within the accuracy
of the method, in the solution we need to retain the terms $\propto
\alpha^n[\ln(\Lambda/\epsilon)]^n$ while same-order terms with a lower
exponent of the logarithmic factor should be disregarded.
 This allows
for a substantial simplification. 
We proceed by solving Eqs.~(\ref{transamp}) and (\ref{ramp}) at higher
energy (far from the resonance), where the reflection from the
compound scatterer is almost perfect. In this case, we approximate
$|r_{L,R}(- E)| \approx 1$. It is possible to see that in this
case the renormalization of the tunnel amplitudes $t_{1,2}$ of each
constituent of our compound scatterer occurs {\it separate} from each
other, $d \ln t_{1,2}/d \ln \epsilon = \alpha_{L,R}/2$.  This
renormalization can be incorporated into the energy dependence of the
effective level widths, $\Gamma_{R,L}(\epsilon)=
\Gamma_{R,L}(\epsilon/\Lambda)^{\alpha_{R,L}}$.  The result
for $|t(\epsilon)|^2$ thus reads
\begin{equation}
|t(\epsilon)|^2 = 
\frac{\Gamma_L(\epsilon) \Gamma_R(\epsilon)}
{(\epsilon - \Delta)^2 +[\Gamma_L(\epsilon) +\Gamma_R(\epsilon)]^2/4}.
\label{high_en}
\end{equation}

The above approximation of the integrand in Eq.~(\ref{transamp})
becomes invalid at lower energies, where the transmission coefficient
may become of the order of unity. The energy scale
$\tilde\epsilon$ at which this occurs can be evaluated from
Eq.~(\ref{high_en}), and is given by the solution of equation $2
\tilde \epsilon =\Gamma_L({\tilde \epsilon})+\Gamma_R({\tilde
  \epsilon})$.  In the simplest case of
$\alpha_L=\alpha_R\equiv\alpha\ll 1$, it is
$2\tilde\epsilon=(\Gamma_L+\Gamma_R)((\Gamma_L+\Gamma_R)/2\Lambda)^\alpha$.
At energies below $\tilde \epsilon$, the reflection amplitudes in the
integrand can be approximated as $r(\epsilon') \approx r(\epsilon)$.
Under this assumption, we immediately recover Eq.(\ref{diffform}). At
$|\epsilon|< \tilde\epsilon$, its solution yields
\begin{equation}
|t(\epsilon)|^2 = 
\frac{\tilde\Gamma_L(\epsilon)\tilde\Gamma_R(\epsilon)}{(\epsilon - \Delta)^2
+\tilde\Gamma_L(\epsilon) \tilde \Gamma_R(\epsilon)
+[{\Gamma_L(\tilde \epsilon)} -{\Gamma_R(\tilde\epsilon)}]^2/4}
\label{lowenergy}
\end{equation}
with 
\begin{equation}
\tilde \Gamma_{L,R}(\epsilon)=\Gamma_{L,R}(\tilde \epsilon)
\left( {|\epsilon|}/{\tilde\epsilon}\right)^{\frac{\alpha_R+\alpha_L}{2}}.
\label{renormgamma1}
\end{equation}
Relations~(\ref{lowenergy}) and (\ref{renormgamma1}) determine the full
crossover function for the resonant tunneling between the interacting
1d electron systems, if $\tilde\epsilon\gtrsim |\Delta|$.

In the opposite case of a resonance distant from the Fermi level,
$|\Delta|\gtrsim\tilde\epsilon$, 
we shall change the approximation at $\epsilon = |\Delta|$.
The answer is thus given by the equations (\ref{lowenergy}),  
(\ref{renormgamma1}) with $\tilde\epsilon$ being {\it replaced} by $|\Delta|$.

A typical energy dependence of the transmission probability is
sketched in the insert of Fig.~1. It combines an overall Lorentz-like
shape with a sharp dip at the Fermi level. Note, that the transmission
is not suppressed at low energies for a perfectly symmetric barrier
with the resonant level tuned to coincide with the Fermi level, in
full agreement with Ref.~\cite{KF}.

It may seem that the numerical factors in the definition of the
crossover energy $\tilde \epsilon$ and in the condition $|\Delta|=
\tilde \epsilon$ of the crossover between low energy cut-offs are
chosen in a rather arbitrary fashion. Indeed, these two definitions
could contain any other numerical factors of the order of $1$.  The
point is that fixing the numerical factors with a greater precision
would exceed the accuracy of our renormalization method. In other
words, the energy dependence of $\Gamma$ in all above relations is
assumed to be very slow. It is this slow dependence that, in the limit
$\alpha \ll 1$, gives us the luxury of arbitrary choice of those
numerical factors.

To present quantitative conclusions, we discuss the linear conductance
$G(T)$ in the case of $\alpha_R=\alpha_L\equiv \alpha$. Within the
Landauer formalism, the conductance is given by
\begin{equation}
G(T)=G_Q 
\int_{-\infty}^{\infty} \frac{d\epsilon}{4 T \cosh^2(\epsilon/2T)} 
|t(\epsilon)|^2,
\label{integral}
\end{equation}
where the conductance quantum unit for one fermion mode is
$G_Q=e^2/2\pi\hbar$.  The results strongly depend on the ratio of
$\Gamma_R$ and $\Gamma_L$.  We will characterize this ratio by the
asymmetry parameter $\beta \equiv
|\Gamma_L-\Gamma_R|/(\Gamma_R+\Gamma_L)$ which ranges from $0$ to $1$
and does not depend on energy, provided that $\alpha_R=\alpha_L$.  To
emphasize the effect of interaction, let us recall that in the case of
free electrons one finds $G(T) \propto 1/T$ at temperatures $T\gg
\Gamma,\Delta$; in the limit $T\to 0$, the conductance saturates at a
finite value, which reaches $(1-\beta^2)G_Q$ if the Fermi level is
tuned to the resonance ($\Delta=0$). Interaction changes this picture
noticeably.  Let us start the discussion with the case $\Delta=0$.  At
high temperatures, $T\gtrsim\tilde\epsilon$, the conductance can be
estimated as
\begin{equation}
{G_{\Delta=0}(T)}/{G_Q} =  
[{\pi (1-\beta^2)}/{4}]\left({T}/{\tilde\epsilon}\right)^{\alpha-1} 
\simeq {\Gamma(T)}/{T}.
\nonumber
\end{equation}
The temperature dependence can be thus ascribed to interaction-induced
power-law temperature dependence of $\Gamma$. Furthermore, the
low-temperature behavior differs strikingly for symmetric ($\beta=0$)
and asymmetric ($\beta \ne 0$) resonance. For symmetric resonance, the
conductance saturates at the ideal value of $G_Q$. For $\beta\neq 0$
the conductance reaches at $T \approx \tilde \epsilon$ its maximum
value, which is smaller than $(1-\beta^2)G_Q$, and drops to zero with the
further decrease of temperature,
\begin{equation}
{G_{\Delta=0}(T)}/{G_Q} =
\left({1}/{\beta^2}-1\right)
\left({T}/{\tilde \epsilon}\right)^{2 \alpha},\quad
T\lesssim\tilde\epsilon.
\label{singlebarrier}
\end{equation}
The temperature exponents at $T\lesssim\tilde\epsilon$ in both cases
agree with those obtained in Refs.~\cite{KF,KF1}. The
exponent at $\beta\neq 0$ is the same as for a single tunnel barrier
interrupting the 1d channel. It indicates that at low energies the
electrons get over the compound scatterer in a single quantum
transition. The high-temperature exponent arises from the separate
renormalization of each barrier, which signals the
``sequential mechanism'' of tunneling: the electron tunnels across one
barrier first and waits a while ($\simeq \hbar/\Gamma(\epsilon)$)
before tunneling across another one.  It is important to recognize
though that no energy relaxation or decoherence takes place during
this waiting time.  This is especially clear from our calculation
based on the Landauer formula: there are no inelastic processes
included which could provide for the relaxation or decoherence.

The increase of $\Delta$ leads to a decrease of the conductance. For
non-interacting electrons, the conductance stays at a level of the
order of its maximal value, $G_{\Delta=0}$ for $\Delta$ less than
$\Gamma_L+\Gamma_R$, which determines the width of the resonance in
$G(\Delta)$ at $T\lesssim\Gamma_L+\Gamma_R$. At higher temperatures,
the effective resonance width is $w\simeq T$. Let us discuss now the
temperature dependence $w(T)$ and the shape of the resonance
$G(\Delta)$ at fixed temperature in the presence of interaction. For
$T\gg\tilde\epsilon$, the width $w\simeq T$ does not reveal any
anomalous exponent.  The shape of the resonance in this regime is
mainly determined by the thermal-activated exponential contribution
$G(\Delta)\simeq \exp(-|\Delta|/T) \Gamma(T)/T$ in
Eq.~(\ref{integral}). However, at large $\Delta\gg w$ the power-law
"cotunneling" tail
\[
G_{\rm tail}(\Delta)
=G_Q(1-\beta^2) \left({T}/{\tilde \epsilon}\right)^{2\alpha} 
{\tilde \epsilon^2 }/{\Delta^2},
\]
replaces that exponential dependence~\cite{Furusaki}. The crossover
occurs at $\Delta \simeq T \ln(G_Q/G_{\Delta=0})$ and corresponds to
the conductance $G_{\rm cross} \simeq G_{\Delta=0}^2/G_{Q}$, this
being much smaller than $G_{\Delta=0}$.

At $T\ll\tilde\epsilon$ the apparent width of the non-symmetric
resonance saturates at $w\simeq\tilde\epsilon$. The conductance
thus drops uniformly at any $\Delta$ following the power law
(\ref{singlebarrier}). The symmetric resonance presents an exception.
In this case, the width shrinks with the decreasing temperature,
\[
w(T)\simeq \left({T}/{\tilde \epsilon}\right)^{\alpha}\tilde\epsilon,
\]
and $G(T,\Delta)$ acquires the scaling form,
$G(T,\Delta)=G_Q/\{1+[\Delta/w(T)]^2\}$, in agreement with
Ref.~\cite{KF}.

We further illustrate our results by a numerical evaluation of
Eq.~(\ref{integral}), see Figs~1--2. For this calculation, we choose
$\alpha=0.2$. By virtue of our approach, the relative accuracy of the
results is expected to be of the order of $\alpha$. The dependence
$G(T)$ is not monotonic, and in the limit $T\to 0$ the conductance
drops to zero at any $\beta\neq 0$, although for small $\beta$ this is
noticeable only at very low temperatures ( Fig.~1). The temperature
dependence $w(T)$ of the width of the resonance $G(\Delta)$ is shown
in the left panel of Fig.~2. If $\beta\neq 0$, this dependence
saturates at some value $w(0)\neq 0$. 

\begin{figure}
\epsfxsize=1.5in 
\centerline{\epsfbox{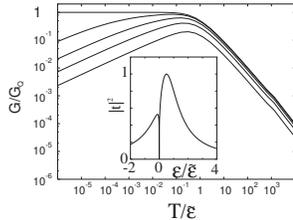}}
\caption{\label{fig1} Temperature dependence of resonant ($\Delta=0$) tunneling
conductance. The asymmetry parameter $\beta=0$ (top curve), $0.2$,
$0.4$, $0.6$, and $0.8$ (bottom curve).
For symmetric resonance ($\beta=0$), the conductance saturates at $T=0$. 
Inset: The typical energy dependence of transmission coefficient. }
\end{figure}
\begin{figure}
\epsfxsize=1.5in
\centerline{\epsfbox{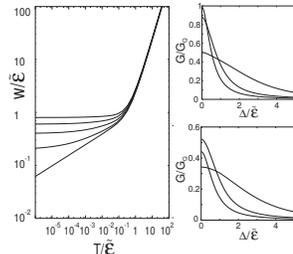}}
\caption{\label{fig2} Left: Half width at half maximum $w$ vs.
temperature $T$ for the
values of asymmetry parameter $\beta=0,0.2,0.4,0.6,0.8$ (bottom to
top curve).  With the decreasing temperature, the half width
saturates for a non-symmetric resonance, and continuously decreases
for the symmetric one. Right: The conductance dependence on the
position of the resonant level with respect to the Fermi level,
$G(\Delta)$, for symmetric (top) and non-symmetric with $\beta=0.5$
(bottom)
resonances at three temperatures $T/\tilde{\epsilon} = 0.04,0.2,1$. }
\end{figure}

The differences and similarities of symmetric and non-symmetric
resonances are further illustrated in the right panels of Fig.~2. The
three pairs of line shapes there correspond to "high", "medium", and
"low" temperatures, respectively. The two high-temperature curves (the
smallest values of $G_{\Delta=0}$) are hardly distinguishable from
each other, and correspond to the resonance width $w\simeq T$. Both
medium-temperature curves show a more narrow resonant peak with
increased conductivity $G_{\Delta=0}$, and are still similar to each
other, apart from the scale. The real difference becomes visible for
the low-temperature curves.  In the case of non-symmetric resonance,
the low-temperature curve is just reduced in height with no noticeable
change of the shape. This is in contrast to the symmetric resonance,
where the resonance peak gets taller and thinner.  The symmetric
resonance does not seem to be very realistic since any randomness in
the barriers and/or in nanowire would cause asymmetry.  Provided
symmetry is achieved, the symmetric resonance can be easily identified
by its ideal conductance $G_Q$.

To conclude, we have investigated the transmission resonances of
interacting electrons in 1d wires. For a weak electron-electron
interaction the transmission can be considered as an elastic process,
which allowed us to build a comprehensive theory of the resonances,
valid in a broad range of temperature and parameters of the resonant
level.  The temperature dependence of the maximum conductance
in general is not monotonic, and reveals important differences between
symmetric and non-symmetric resonances. The obtained quantitative
results present a comprehensive and consistent picture of the effect.
It assures us in the qualitative validity of the picture at an
arbitrary interaction strength. Although we are not able to come up
with an explicit expression for the crossover function $G(T)$ in this
case, such function, with known high- and low-temperature asymptotes,
does exist by virtue of the renormalizability.

We acknowledge a useful communication with D. Polyakov and I. Gornyi
\cite{Polyakov} which enabled us to present Eqs.~(\ref{transamp}) and
(\ref{ramp}) in a correct and comprehensible way.  One of the authors
appreciates many stimulating discussions with M. Grifoni and M.
Thorwat. This work has been initiated in the framework of 2001 Aspen
Summer Program and was partially done during a workshop at Institute
of Theoretical Physics, UCSB.  The hospitality of both organizations
is gladly acknowledged.  This research was sponsored by the NSF Grants
DMR 97-31756, DMR 02-37296, EIA 02-10736, and the grants of FOM.

\end{multicols}
\end{document}